\documentclass[preprint,showpacs,preprintnumbers,amsmath,amssymb]{revtex4}

\usepackage{graphicx}
\usepackage{dcolumn}
\usepackage{bm}
\usepackage{mathrsfs}

\begin{document}


\title{Cylindrical Invisibility Cloak with Simplified Material Parameters is Inherently Visible}

\author{Min Yan}
\author{Zhichao Ruan}
\author{Min Qiu}
\email{min@kth.se}
\affiliation{%
Department of Microelectronics and Applied Physics, \\
Royal Institute of Technology, Electrum 229, 16440 Kista, Sweden
}%

\date{\today}

\begin{abstract}
It was proposed that perfect invisibility cloaks can be constructed
for hiding objects from electromagnetic illumination (Pendry et al., 
Science 312, p.~1780). The cylindrical cloaks experimentally demonstrated (Schurig et
al., Science 314, p.~997) and theoretically proposed (Cai et al., Nat. Photon.
1, p.~224) have however simplified material parameters in order to
facilitate easier realization as well as to avoid infinities in
optical constants. Here we show that the cylindrical cloaks with
simplified material parameters inherently allow the zero$^{th}$-order
cylindrical wave to pass through the cloak as if the cloak is made of a homogeneous isotropic
medium, and thus visible. To all high-order cylindrical waves, our
numerical simulation suggests that the simplified cloak
inherits some properties of the ideal cloak, but finite scatterings exist.
\end{abstract}

\maketitle

Recently there has been an increase of interest in realizing
invisibility cloaks \cite{Greenleaf:StaticCloaking,
  Greenleaf:CloakingAllFreq, Milton:TransformationInvariantForm,
  Leonhardt:conformal, Pendry:cloak, Miller:perfectCloaking,
  Alu:transparency, Chen:CloakConstraints}.  In particular, the coordinate
transformation method proposed in \cite{Pendry:cloak} is noticed to be
especially powerful for designing invisibility cloaks that can in
principle completely shield enclosed
objects from electromagnetic (EM) illumination and at the
same time cause zero disturbance to the foreign EM field. In 2D case,
the cylindrical cloak obtained through
such technique has anisotropic, spatially varying optical constants.
In addition, some of the material parameters have
infinite values at the interior surface of the cloak. To
facilitate experimental realization at microwave frequencies, Schurig
et al. have simplified the material properties such that only one
material parameter is gradient and the requirement on infinite material constant
is lifted \cite{Schurig:cloak}. The authors claimed that, in
comparison to the ideal cloak, the simplified cloak maintains the power-flow bending
with the penalty of nonzero reflectance at the outer
interface. Similar simplification on a
cylindrical cloak at optical frequencies is also employed in
Ref. \cite{Cai:cloakTM}. In this paper we provide a systematic
theoretical study on simplified cylindrical cloaks. It is found that
the bare device constructed using the simplified medium fails to be
invisible. In addition, the device does not possess a spatial region that is
completely in isolation from the outside world
electromagnetically. Hence, perfect hiding with such a simplfied cloak
is not possible.

Consider a cylindrical cloak with its inner and outer boudaries
positioned at $r=a$ and $r=b$ respectively. Both domains inside and
outside the cloak are air. The structure is in general a three-layered cylindrical
scatterer. We refer to the layers from inside to outside as layer 1, 2 and
3, respectively. 
Our analysis, similar to all previous works, is on normal
incidence. That is, the $\mathbf{k}$ vector is perpendicular to the
cloak cylinder axis. The EM
wave is assumed to have TE polarization (i.e. electric field only
exists in $z$ direction). The TM polarization case
can be derived by making $E\rightarrow -H$, $\varepsilon\rightarrow \mu$,
and $\mu\rightarrow \varepsilon$ substitutions. By default, we choose the cloak's
cylindrical coordinate as the global coordinate. In a homogeneous
material region (i.e. cloak interior and exterior) the
general solution is expressible in Bessel functions. Within the cloak
medium, the general wave equation that governs the $E_z$ field can
be written as
\begin{equation}
\frac{1}{r}\left[\frac{\partial}{\partial
    r}\left(\frac{r}{\mu_\theta}\frac{\partial E_z}{\partial
      r}\right)\right]+\frac{1}{r^2}\frac{\partial}{\partial\theta}\left( \frac{1}{\mu_r}\frac{\partial E_z}{\partial \theta}\right)+k_0^2\varepsilon_z E_z=0,
\label{eq:waveEqGeneral}
\end{equation}
where $k_0$ is the free-space wave number, $\mu_\theta$, $\mu_r$ and
$\varepsilon_r$ are polarization-dependent permeability/permitivity profiles of the cloak. The time dependence $\exp(i\omega t)$
 has been used for deriving Eq. \ref{eq:waveEqGeneral}.

Ideally, the cloak is designed to compress all
fields within a cylindrical air region $r<b$ into the cylindrical
annular region $a<r<b$. A corresponding coordinate transformation
leads to a set of anisotropic and \emph{spatially variant} material constants in the cloak shell,
as described in \cite{Schurig:cloak}.
Such an annular cylinder can indeed provide perfect invisibility cloaking
\cite{Ruan:cloak}, but it requires infinite values of optical
constants at the cloak's inner boundary. To circumvent the fabrication
difficulty, the simplified parameters were used \cite{Schurig:cloak}. They are in the form of
\begin{eqnarray}
&\mu_r=\left(\frac{r-a}{r}\right)^2, \nonumber\\
&\mu_\theta=1,\label{eq:simplified}
\\
&\varepsilon_z=\left(\frac{b}{b-a}\right)^2.
\nonumber
\end{eqnarray}
To achieve cloaking for TM waves, the same set of material
parameters, but for $\varepsilon_r$, $\varepsilon_\theta$ and
$\mu_z$, is used \cite{Cai:cloakTM}.
However, we notice that the precedure for simplification of the
material parameters adopted in \cite{Schurig:cloak} is questionable, as
the derivation has assumed beforehand that $\mu_\theta$ is a constant. It is
obvious that the invariant $\mu_\theta$ can be taken out of the
differential operator in Eq. \ref{eq:waveEqGeneral}. Therefore wave
behavior within the cloak shell is altered as compared to that in an
ideal cloak.

Since the material parameters in Eq. \ref{eq:simplified} are
 azimuthally invariant (which is also true for the ideal parameter
 set), we can use the variable separation
 $E_z=\Psi(r)\Theta(\theta)$. Eq. \ref{eq:waveEqGeneral} can then
 be decomposed into
\begin{eqnarray}
&&\frac{d^2\Theta}{d\theta^2}+m^2\Theta=0,\label{eq:waveEqTheta}\\
&&\frac{d}{dr}\left(\frac{r}{\mu_\theta}\frac{d\Psi}{dr}\right)+k_0^2r\varepsilon_z\Psi
-m^2\frac{1}{r\mu_r}\Psi=0,
\label{eq:waveEqPsi}
\end{eqnarray}
where $m$ is an integer. The solution to Eq. \ref{eq:waveEqTheta} is
$\exp(im\theta)$. Equation \ref{eq:waveEqPsi} is a
second-order homogeneous differential equation. Two independent
solutions are expected. At this moment, we assume the solution to
Eq. \ref{eq:waveEqPsi}, for a fixed $m$, can be written in general as
$\mathscr{A}_mQ_m+\mathscr{B}_mR_m$, where $\mathscr{A}_m$ and $\mathscr{B}_m$
are constants. $Q_m$ and $R_m$ are functions of
$r$. Now valid field solutions in three layers (denoted by superscripts) can be
described as
\begin{eqnarray}
E_z^1&=&\sum_m\mathscr{A}_m^1J_m(k_0r)\exp(im\theta),\\
E_z^2&=&\sum_m\{\mathscr{A}_m^2Q_m+\mathscr{B}_m^2R_m\}\exp(im\theta),\\
E_z^3&=&\sum_m\{\mathscr{A}_m^3J_m(k_0r)+\mathscr{B}_m^3H_m^{(2)}(k_0r)\}\exp(im\theta).
\label{eq:fieldL}
\end{eqnarray}
$H_m^{(2)}$ is the Hankel function of the second kind, which
represents outward-travelling cylindrical wave. The $J_m$ and $H_m^{(2)}$ terms in
the 3rd layer are physically in correspondence to the incident and scattered waves,
respectively. Hence, the scattering problem becomes to solve for,
most importantly, $\mathscr{A}_m^1$ (transmitted field) and $\mathscr{B}_m^3$ (scattered
field) subject to a given incidence $\mathscr{A}_m^3$ 
\cite{note:A3}. Comparatively $\mathscr{A}_m^2$ and $\mathscr{B}_m^2$ are
physically less interesting. The coefficients are solved by matching
the tangential fields ($E_z$ and $H_\theta$) at the layer interfaces. Due to the
orthogonality of the function $\exp(im\theta)$, the cylindrical waves
in different orders decouple. We hence can examine the transmission
and scattering of the cloak for each individual order number
$m$.

By substituting the simplified material parameters into
Eq. \ref{eq:waveEqPsi}, we obtain
\begin{equation}
 (r - a)^2 \frac{{d^2 \Psi }}{{dr^2 }} + \frac{(r - a)^2}{r}\frac{{d\Psi }}{{dr}} + \left[(r - a)^2\left(\frac{b}{{b - a}}\right)^2 k_0^2  - m^2 \right]\Psi  = 0. \\
\label{eq:waveEqReducedPsi}
\end{equation}
Equation \ref{eq:waveEqReducedPsi} has two non-essential
singularities at $r=0$ and $r=a$ for $m\neq 0$. It is worthwhile to mention that, with the ideal parameters,
Eq. \ref{eq:waveEqPsi} can be written instead as
\begin{equation}
 (r - a)^2 \frac{{d^2 \Psi }}{{dr^2 }} + (r - a)\frac{{d\Psi }}{{dr}} + \left[(r - a)^2\left(\frac{b}{{b - a}}\right)^2 k_0^2  - m^2 \right]\Psi  = 0.
\label{eq:waveEqIdealPsi}
\end{equation}

When $m=0$, Eq \ref{eq:waveEqReducedPsi} can be further simplified to
\begin{equation}
r^2\frac{d^2 \Psi}{d r^2}+r\frac{d \Psi}{d
  r}+r^2k_0^2\left(\frac{b}{b-a}\right)^2 \Psi=0.
\label{eq:waveEqReducedPsiJ0}
\end{equation}
This is the zero$^{th}$-order Bessel differential equation. Its 
non-essential singularity remains at $r=0$. Equation
\ref{eq:waveEqReducedPsiJ0} suggests that an incoming
zero$^{th}$-order cylindrical wave would effectively see the
simplified cloak as a homogeneous isotropic medium whose effective
refractive index is $n_\mathrm{eff}=\frac{b}{b-a}$. Its transmission
through the cloak shell is therefore determined by the etalon effect of
the finite medium. 

When $m\neq 0$, the wave solution is governed by
Eq. \ref{eq:waveEqReducedPsi}. By comparing
Eqs. \ref{eq:waveEqIdealPsi} and \ref{eq:waveEqReducedPsi}, we see
that at radial positions $r>>a$, Eq. \ref{eq:waveEqReducedPsi}
asympototically resembles Eq. \ref{eq:waveEqIdealPsi} due to
$r-a\approx r$. This indicates that all high-order cylindrical waves tend to
behave similarly in both media at $r>>a$ positions, and hence the
importance of parameter $b$, at which the cloak medium is
truncated. In the following, we will derive the scattering coefficient
$s_m$ subject to individual cylindrical wave incidences. The
scattering coefficient in each cylindrical order is defined as
$s_m=\left|\mathscr{B}_m^3/\mathscr{A}_m^3\right|$. Analytic derivation of $s_m$
can be done if two solutions to Eq. \ref{eq:waveEqReducedPsi}
(i.e. $Q_m$ and $R_m$) are known in closed form. However, despite the
analogueness to Eq. \ref{eq:waveEqIdealPsi},
Eq. \ref{eq:waveEqReducedPsi} fails the analytic Frobenius method \cite{Arfken:mathPhy}. 
Here we tackle the problem through the finite-element method. The
field outside the cloak is computed numerically, and then is used for
deriving coefficients $\mathscr{A}_m^3$ and $\mathscr{B}_m^3$ through a fitting procedure. 
$s_m$ is known in turn.  Solutions with different azimuthal orders are obtained by
varying the azimuthal dependence of a circular current source outside
the cloak. The
scattering problem is numerically manageable since functionals
$\varepsilon_z\mu_\theta$ and $\frac{\mu_\theta}{\mu_r}$ in
Eq. \ref{eq:waveEqPsi}, unlike in the ideal cloak case, are both finite and do not possess any
removable singularity for the simplified medium.
The commercial software COMSOL is deployed to carry out
calculations. For our
case study, we fix $a=0.1$m and operating
frequency $f=2$GHz. The performance of the cloak is examined as $b$ is
increased from $0.2$m. Similar parameters are also found in
\cite{cummer:036621}.

\begin{figure}[h]
\centering
\includegraphics[width=12cm]{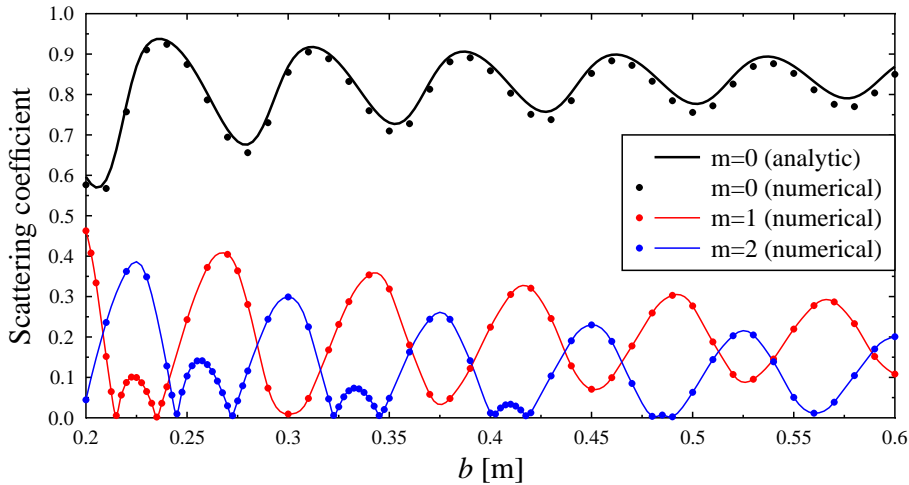}
\caption{Variation of the scattering coefficients, examined in each
  cylindrical wave order, as a function of $b$. For $m=1,2$, the curves are fitted using
  Savitzky-Golay smoothing filter according to numerically derived
  data points (dots).}
\label{fig:scat}
\end{figure}

The scattering coefficients in different azimuthal orders as a
function of $b$ are shown in Fig. \ref{fig:scat}. As expected,
the zero$^{th}$-order scattering coefficient is quite
distinct from others due to the different governing wave
equation. In fact, analytic solution exists when $m=0$, as field
in the cloak medium are Bessel functions \cite{Ruan:cloak}. The excellent agreement
between the numerical and analytic results for zero$^{th}$-order
scattering coefficient confirms the validity and accuracy of our
approach. Using the analytic technique, the zero$^{th}$-order
scattering coefficient is found to converge to 0.867 with respect to
$b$. Despite that the effective index of the cloak approaches to 1 as
$b$ increases, the phase variation of the zero$^{th}$-order wave
within the cloak medium is increasing, as
$k_0\frac{b}{b-a}(b-a)=k_0b$. This explains why the scattering
coefficient converges to a value other than 0. The existence of
zero$^{th}$-order scattering coefficient effectively disqualifies the
cloak to be completely invisible.

Compared to the zero$^{th}$-order scattering coefficient, the
high-order scattering coefficients (only those for $m=1,2$ are shown
in Fig. \ref{fig:scat}) are noticed to be in a similar
oscillatory fashion, and in general much smaller. The
scattering coefficient tends to converge to a value closer to zero
when the order number $m$ increases. Over certain ranges of $b$ value
(e.g. around $b=0.225$m for $m=1$), the computed
$\mathscr{B}_m^3$ changes sign and hence the
resulted scattering coefficient seems to be flipped from a negative
value. We should attribute the relatively small high-order scattering
coeffcients to the cloak's partial inherence of the ideal cloak based
on coordinate transformation \cite{note:comparison}. However, our numerical
result shows that the high-order scattering coefficients do not
converge to zero even when the cloak wall is very thick.

\begin{figure}[b]
\centering
\includegraphics[width=12cm]{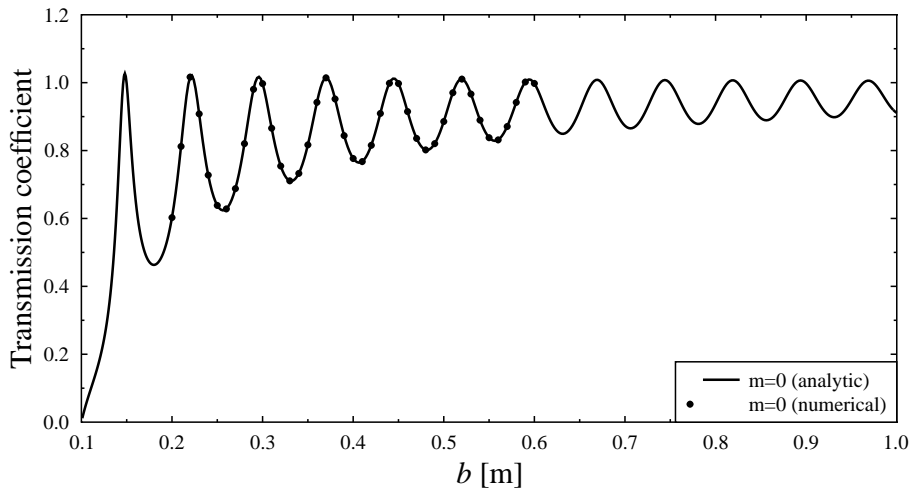}
\caption{The zero$^{th}$-order transmission coefficient.}
\label{fig:trans}
\end{figure}

Besides the requirement of zero scattering (invisibility), a device
also need to possess a spatial region which is in complete EM
isolation from the exterior world in order to be an invisibility
cloak. Therefore it is meaningful to know how much field penetrates
into the simplified cloak subject to a foreign EM illumination. Again,
the problem is studied by examining the individual cylindrical wave
components separately. The transmission coefficient, defined as
$t_m=\left|\mathscr{A}_m^1/\mathscr{A}_m^3\right|$, is used to characterize
the field transmission. When $m=0$, the amount of field transmitted into the cloak interior can be
analytically derived, which is shown in
Fig. \ref{fig:trans} as a function of $b$. The transmission is noticed to be
oscillatory, and converging to 1 as $b$ increases. Numerical
calculation is also superimposed for validation. When $m\neq 0$, the
FEM calculations show that the field inside the cloak is almost
zero. The corresponding transmission coefficients are exclusively
smaller than 0.005, hence are not plotted in
Fig. \ref{fig:trans}. This indicates that the contour $r=a$
provides an insulation between its enclosed domain and the exterior
domain, but only for all high-order cylindrical waves. Therefore, any objects
placed inside the cloak are exposed to the zero$^{th}$-order
cylindrical wave component. Reversely, the zero$^{th}$-order wave
component of an EM source placed within the cloak (or scattered wave
by objects inside the cloak) will transmit out.
As a result, objects enclosed by a simplified cloak is
sensible by a foreign detection unit.

\begin{figure}[h]
\centering
\includegraphics[width=12.8cm]{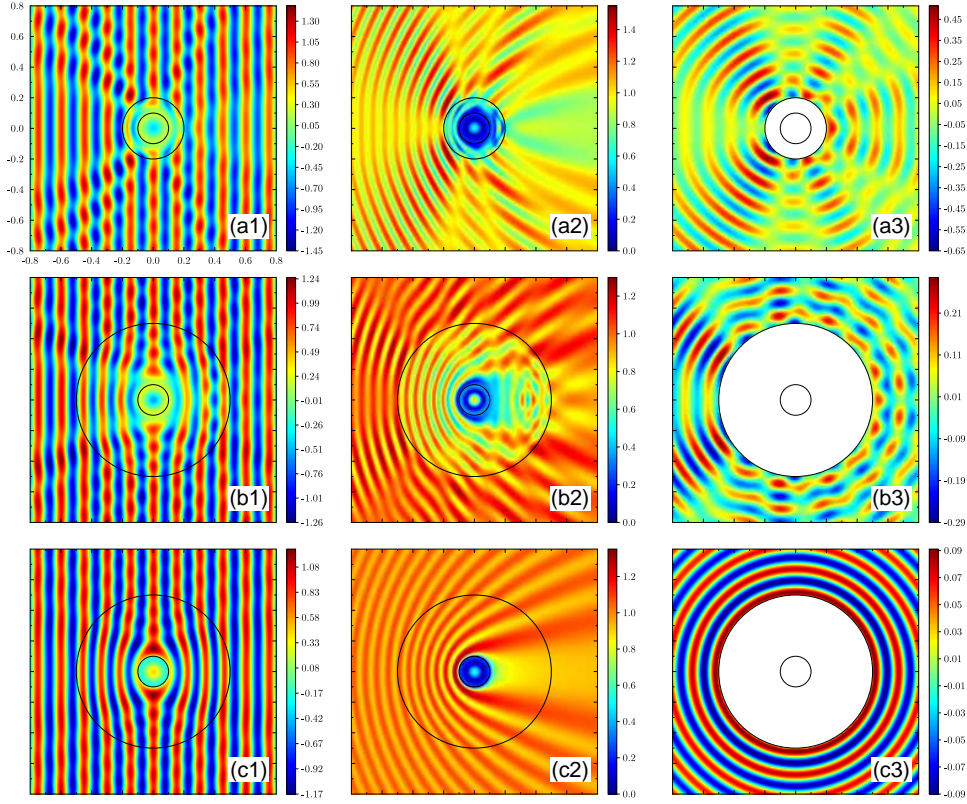}
\caption{(a1)-(a3) $E_z$ snapshot, $E_z$ norm, and scattered $E_z$
  field, respectively, for a simplified cloak with $a=0.1$m and
  $b=0.2$m. (b1)-(b3) Same fields but for a simplified cloak with
  $a=0.1$m and $b=0.5$m. (c1)-(c3) Same fields but for a near-to-ideal
  cloak with $a=0.101$m and $b=0.5$m.}
\label{fig:fields}
\end{figure}

Next we numerically demonstrate (in COMSOL) scattering by the
simplified cloaks with a plane-wave incidence. The incident plane wave
travels from left to right and has the amplitude 
of 1. When $b=0.2$m, the $E_z$ snapshot, $E_z$ norm, and the scattered
$E_z$ snapshot are plotted in Fig. \ref{fig:fields}(a1)-(a3), respectively. It is
noticed that the amplitude of the scattered field is about one half of the
incident field. From the scattered field distribution, high-order
Bessel terms constitute a significant portion. Scattering aside, the
field inside the cloak shell is seen to have only zero$^{th}$-order
Bessel term. We then increase $b$ to 0.5m,
and the corresponding numerical results are plotted in
Fig. \ref{fig:fields}(b1)-(b3). Compared to the previous case,
scattered field is reduced roughly by half in amplitude, and
is now dominated by the zero$^{th}$-order Bessel term. The $E_z$ norm is
closer to be uniform outside the cloak, indicating better
invisibility. The overall smaller scattering as well as the dominance
of the zero$^{th}$-order scattering for the second cloak agree well with
our derviation of the scattering coefficients in Fig. \ref{fig:scat}. 
When $b$ is changed from 0.2m to 0.5m, the transmitted $E_z$ field at
the center of the cloak increases from 0.6032 to 0.8855 in
norm, also in agreement with Fig. \ref{fig:trans}.

For completeness, we also show the scattering by a
near-to-ideal cloak in Fig. \ref{fig:fields}(c1)-(c3). The material
parameters of the cloak are defined by $\mu_r=\frac{r-a}{r}$,
$\mu_\theta=\frac{r}{r-a}$, and $\varepsilon_z=\left(\frac{b}{b-a}\right)^2\frac{r-a}{r}$, with
$a=0.1$m and $b=0.5$m. To avoid
the $r=0.1$m critical contour, we let the inner boundary of the cloak be
positioned at $r=0.101$m. In \cite{Ruan:cloak} we have confirmed
analytically that the performance of an ideal cloak is extremely
sensitive to the position of the cloak's inner surface. In particular,
the zero$^{th}$-order cylindrical wave will experience considerable
reflection and transmission as it meets the cloak shell. This is
confirmed in Fig. \ref{fig:fields}(c1)-(c3). The scattered field is
almost purely the zero$^{th}$-order cylindrical wave. Even at such a
small mis-location of the inner boundary, the amount of field leaking
into the cloak is handsome, valued at 0.5466 in norm. By comparing
panels (c1)-(c3) and (b1)-(b3) in Fig. \ref{fig:fields}, we see that
the improved cloak made from simplified medium inherits some merits of
the near-to-ideal cloak, but with overall higher scattering, especially
in the monopole wave component.

In conclusion, we have theoretically studied the cylindrical cloaks
with simplified material parameters. Such a simplified cloak is shown to inherit
some properties of an ideal cloak based on coordinate transformation
of Maxwell equations. However the penalty of using the simplified
cloak is more than just nonzero reflectance at the cloak boundary. The
monopole component of the incoming wave will always experience
relatively high scattering. High-order cylindrical waves also
experience finite (although smaller) scattering even when the cloak's
wall is kept very thick, as suggested by our numerical simulation.
Besides that the device itself is visible, it cannot
completely shield EM field, due to penetration of the monopole
field component. Hence cloaking of objects will not be perfect.
Lastly, considering that the
monopole field treats a simplified cloak as a conventional
glass tube, detection of any object placed within the cloak can be
greatly enhanced by using a EM source with the maximum energy on its
monopole component.

\noindent{\bf Acknowledgement:} This work is supported by the Swedish Foundation for Strategic
Research (SSF) through the INGVAR program, the SSF Strategic Research
Center in Photonics, and the Swedish Research Council (VR).


\begin{thebibliography}{15}
\expandafter\ifx\csname natexlab\endcsname\relax\def\natexlab#1{#1}\fi
\expandafter\ifx\csname bibnamefont\endcsname\relax
  \def\bibnamefont#1{#1}\fi
\expandafter\ifx\csname bibfnamefont\endcsname\relax
  \def\bibfnamefont#1{#1}\fi
\expandafter\ifx\csname citenamefont\endcsname\relax
  \def\citenamefont#1{#1}\fi
\expandafter\ifx\csname url\endcsname\relax
  \def\url#1{\texttt{#1}}\fi
\expandafter\ifx\csname urlprefix\endcsname\relax\def\urlprefix{URL }\fi
\providecommand{\bibinfo}[2]{#2}
\providecommand{\eprint}[2][]{\url{#2}}

\bibitem[{\citenamefont{Greenleaf et~al.}(2003)\citenamefont{Greenleaf, Lassas,
  and Uhlmann}}]{Greenleaf:StaticCloaking}
\bibinfo{author}{\bibfnamefont{A.}~\bibnamefont{Greenleaf}},
  \bibinfo{author}{\bibfnamefont{M.}~\bibnamefont{Lassas}}, \bibnamefont{and}
  \bibinfo{author}{\bibfnamefont{G.}~\bibnamefont{Uhlmann}},
  \bibinfo{journal}{Physiol. Meas.} \textbf{\bibinfo{volume}{24}},
  \bibinfo{pages}{413} (\bibinfo{year}{2003}).

\bibitem[{\citenamefont{Al{\`u} and Engheta}(2005)}]{Alu:transparency}
\bibinfo{author}{\bibfnamefont{A.}~\bibnamefont{Al{\`u}}} \bibnamefont{and}
  \bibinfo{author}{\bibfnamefont{N.}~\bibnamefont{Engheta}},
  \bibinfo{journal}{Physical Review E} \textbf{\bibinfo{volume}{72}},
  \bibinfo{pages}{016623} (\bibinfo{year}{2005}).

\bibitem[{\citenamefont{Leonhardt}(2006)}]{Leonhardt:conformal}
\bibinfo{author}{\bibfnamefont{U.}~\bibnamefont{Leonhardt}},
  \bibinfo{journal}{Science} \textbf{\bibinfo{volume}{312}},
  \bibinfo{pages}{1777} (\bibinfo{year}{2006}).

\bibitem[{\citenamefont{Pendry et~al.}(2006)\citenamefont{Pendry, Schurig, and
  Smith}}]{Pendry:cloak}
\bibinfo{author}{\bibfnamefont{J.~B.} \bibnamefont{Pendry}},
  \bibinfo{author}{\bibfnamefont{D.}~\bibnamefont{Schurig}}, \bibnamefont{and}
  \bibinfo{author}{\bibfnamefont{D.~R.} \bibnamefont{Smith}},
  \bibinfo{journal}{Science} \textbf{\bibinfo{volume}{312}},
  \bibinfo{pages}{1780} (\bibinfo{year}{2006}).

\bibitem[{\citenamefont{Miller}(2006)}]{Miller:perfectCloaking}
\bibinfo{author}{\bibfnamefont{D.~A.~B.} \bibnamefont{Miller}},
  \bibinfo{journal}{Optics Express} \textbf{\bibinfo{volume}{14}},
  \bibinfo{pages}{12457} (\bibinfo{year}{2006}).

\bibitem[{\citenamefont{Milton et~al.}(2006)\citenamefont{Milton, Briane, and
  Willis}}]{Milton:TransformationInvariantForm}
\bibinfo{author}{\bibfnamefont{G.~W.} \bibnamefont{Milton}},
  \bibinfo{author}{\bibfnamefont{M.}~\bibnamefont{Briane}}, \bibnamefont{and}
  \bibinfo{author}{\bibfnamefont{J.~R.} \bibnamefont{Willis}},
  \bibinfo{journal}{New. J. Phys.} \textbf{\bibinfo{volume}{8}},
  \bibinfo{pages}{248} (\bibinfo{year}{2006}).

\bibitem[{\citenamefont{Greenleaf et~al.}(2007)\citenamefont{Greenleaf,
  Kurylev, Lassas, and Uhlmann}}]{Greenleaf:CloakingAllFreq}
\bibinfo{author}{\bibfnamefont{A.}~\bibnamefont{Greenleaf}},
  \bibinfo{author}{\bibfnamefont{Y.}~\bibnamefont{Kurylev}},
  \bibinfo{author}{\bibfnamefont{M.}~\bibnamefont{Lassas}}, \bibnamefont{and}
  \bibinfo{author}{\bibfnamefont{G.}~\bibnamefont{Uhlmann}},
  \bibinfo{journal}{Comm. in Math. Phys.} \textbf{\bibinfo{volume}{275}},
  \bibinfo{pages}{749} (\bibinfo{year}{2007}).

\bibitem[{\citenamefont{Chen et~al.}(2007)\citenamefont{Chen, Liang, Yao,
  Jiang, Ma, and Chan}}]{Chen:CloakConstraints}
\bibinfo{author}{\bibfnamefont{H.}~\bibnamefont{Chen}},
  \bibinfo{author}{\bibfnamefont{Z.}~\bibnamefont{Liang}},
  \bibinfo{author}{\bibfnamefont{P.}~\bibnamefont{Yao}},
  \bibinfo{author}{\bibfnamefont{X.}~\bibnamefont{Jiang}},
  \bibinfo{author}{\bibfnamefont{H.}~\bibnamefont{Ma}}, \bibnamefont{and}
  \bibinfo{author}{\bibfnamefont{C.~T.} \bibnamefont{Chan}},
  \bibinfo{journal}{Phys. Rev. B} \textbf{\bibinfo{volume}{76}},
  \bibinfo{pages}{241104(R)} (\bibinfo{year}{2007}).

\bibitem[{\citenamefont{Schurig et~al.}(2006)\citenamefont{Schurig, Mock,
  Justice, Cummer, Pendry, Starr, and Smith}}]{Schurig:cloak}
\bibinfo{author}{\bibfnamefont{D.}~\bibnamefont{Schurig}},
  \bibinfo{author}{\bibfnamefont{J.~J.} \bibnamefont{Mock}},
  \bibinfo{author}{\bibfnamefont{B.~J.} \bibnamefont{Justice}},
  \bibinfo{author}{\bibfnamefont{S.~A.} \bibnamefont{Cummer}},
  \bibinfo{author}{\bibfnamefont{J.~B.} \bibnamefont{Pendry}},
  \bibinfo{author}{\bibfnamefont{A.~F.} \bibnamefont{Starr}}, \bibnamefont{and}
  \bibinfo{author}{\bibfnamefont{D.~R.} \bibnamefont{Smith}},
  \bibinfo{journal}{Science} \textbf{\bibinfo{volume}{314}},
  \bibinfo{pages}{977} (\bibinfo{year}{2006}).

\bibitem[{\citenamefont{Cai et~al.}(2007)\citenamefont{Cai, Chettiar,
  Kildishev, and Shalaev}}]{Cai:cloakTM}
\bibinfo{author}{\bibfnamefont{W.}~\bibnamefont{Cai}},
  \bibinfo{author}{\bibfnamefont{U.~K.} \bibnamefont{Chettiar}},
  \bibinfo{author}{\bibfnamefont{A.~V.} \bibnamefont{Kildishev}},
  \bibnamefont{and} \bibinfo{author}{\bibfnamefont{V.~M.}
  \bibnamefont{Shalaev}}, \bibinfo{journal}{Nat. Photon.}
  \textbf{\bibinfo{volume}{1}}, \bibinfo{pages}{224} (\bibinfo{year}{2007}).

\bibitem[{\citenamefont{Ruan et~al.}(2007)\citenamefont{Ruan, Yan, Neff, and
  Qiu}}]{Ruan:cloak}
\bibinfo{author}{\bibfnamefont{Z.}~\bibnamefont{Ruan}},
  \bibinfo{author}{\bibfnamefont{M.}~\bibnamefont{Yan}},
  \bibinfo{author}{\bibfnamefont{C.~W.} \bibnamefont{Neff}}, \bibnamefont{and}
  \bibinfo{author}{\bibfnamefont{M.}~\bibnamefont{Qiu}},
  \bibinfo{journal}{Phys. Rev. Lett.} \textbf{\bibinfo{volume}{99}},
  \bibinfo{pages}{113903} (\bibinfo{year}{2007}).

\bibitem[{not({\natexlab{a}})}]{note:A3}
\bibinfo{note}{For example, the right-travelling incidence plane wave
  $\exp(-jk_0x)$ can be expanded in Bessel functions, or a generalized Fourier
  series, as $\sum_m{(-i)^mJ_m(k_0r)}\exp(im\theta)$. Therefore
  $\mathscr{A}_m^3$ can be determined beforehand. See D. Felbacq et~al., J.
  Opt. Soc. Am. A {\bf 11}, 2526 (1994).}

\bibitem[{\citenamefont{Arfken}(1970)}]{Arfken:mathPhy}
\bibinfo{author}{\bibfnamefont{G.}~\bibnamefont{Arfken}},
  \emph{\bibinfo{title}{Mathematical Methods for Physicists}}
  (\bibinfo{publisher}{Academic Press}, \bibinfo{year}{1970}),
  chap.~\bibinfo{chapter}{9}, \bibinfo{edition}{2nd} ed.

\bibitem[{\citenamefont{Cummer et~al.}(2006)\citenamefont{Cummer, Popa,
  Schurig, Smith, and Pendry}}]{cummer:036621}
\bibinfo{author}{\bibfnamefont{S.~A.} \bibnamefont{Cummer}},
  \bibinfo{author}{\bibfnamefont{B.-I.} \bibnamefont{Popa}},
  \bibinfo{author}{\bibfnamefont{D.}~\bibnamefont{Schurig}},
  \bibinfo{author}{\bibfnamefont{D.~R.} \bibnamefont{Smith}}, \bibnamefont{and}
  \bibinfo{author}{\bibfnamefont{J.}~\bibnamefont{Pendry}},
  \bibinfo{journal}{Physical Review E} \textbf{\bibinfo{volume}{74}},
  \bibinfo{eid}{036621} (\bibinfo{year}{2006}).

\bibitem[{not({\natexlab{b}})}]{note:comparison}
\bibinfo{note}{For comparison purpose, it should be mentioned that the
  scattering coefficient in any order of an annular cylinder varies between 0
  and 1 as a function of either its refractive index (as geometry is fixed) or
  its outer radius (as material and inner radius are fixed).}

\end{thebibliography}

\end{document}